\documentclass{aa}

\usepackage{graphicx}
\usepackage{txfonts}
\usepackage{natbib}
\bibpunct{(}{)}{;}{a}{}{,}

\newcommand{\Rs}{$ R_{\odot}$}
\newcommand{\de}{$^{\circ}$}
\newcommand{\pB}{\textit{pB}}

\begin{document}

\title{An Empirical 3D Model of the Large-scale Coronal Structure Based on the Distribution of Prominences on the Solar Disk}

\author{Huw Morgan \and S. Rifai Habbal}

\titlerunning{Empirical Model of Coronal Structure Based on Prominences}

\authorrunning{Morgan \& Habbal}

\offprints{Huw Morgan}

\institute{Institute for Astronomy, University of Hawaii, 2680 Woodlawn Drive, Honolulu, HI 96822, USA \\
		\email{hmorgan@ifa.hawaii.edu}}

\date{Recieved September 25, 2006}

%%%%%%%%%%%%%%%%%%%%%%%%%%%%%%%%%%%%%%%%%%%%%%%%%%%%%%%%%%%%%%%%%%%%%%%%%%%%%%%%%%%%%%%%%%%%%%%%%%%%%%%%%%%%%%%%%%%%%%%%%%
\abstract
{Despite the wealth of solar data currently available, the explicit connection between coronal streamers and features on the solar disk remains unresolved.}
{An empirical three-dimensional model, which reproduces the evolution of the large-scale coronal structure starting from the solar surface, is presented. The model is based on the view that the source of the large-scale coronal structure, namely streamers, is a consequence of the evolution of twisted sheet-like structures originating from prominences (or, equivalently, filaments) at the base of the corona.}
 {A 3D model is created whereby high-density sheets are placed above prominences on the solar disk, which evolve and merge with height into a final radial configuration constrained by the oberved position of streamers stalks higher up in the corona. The observational constraints are provided by white light observations from the LASCO/C2 data during the declining phase of solar activity, spanning the end of Carrington Rotation (CR) 2005 and the start of CR 2006, i.e. July-August 2003, and the position of filaments from the corresponding H$\alpha$ synoptic maps of the Paris-Meudon Observatory.}
 {The 3D model thus derived yields a reasonable agreement with the observed large-scale coronal structure, in particular the shape of large helmet streamers.}
 {These results give confidence in the underlying assumption that large helmet streamers can be the result of the convergence of two or more sheet-like structures originating from a distribution of prominences on the solar disk. The model supports the view that streamers, during that time of the solar cycle, are often associated with multiple current sheets.}
%%%%%%%%%%%%%%%%%%%%%%%%%%%%%%%%%%%%%%%%%%%%%%%%%%%%%%%%%%%%%%%%%%%%%%%%%%%%%%%%%%%%%%%%%%%%%%%%%%%%%%%%%%%%%%%%%%%%%%%%%%


\maketitle

\section{INTRODUCTION}
\label{intro}
Establishing connections between observed features in the chromosphere or low in the corona, and structures higher in the corona, is challenging to achieve directly from observations due to the complexity of the corona at low heights. Recent improvements in observations, particularly using spacecraft coronagraphs such as the High Altitude Observatory's white-light coronagraph aboard Skylab, SPARTAN, or the Solar and Heliospheric Observatory (SOHO) Large Angle and Spectrometric Coronagraph Experiment (LASCO) C2 instrument \citep{bru1995}, have allowed sophisticated analyses of coronal structure. \citet{guh1996} created a 3D model of coronal density in relation to the equatorial current sheet during solar minimum. This model was further developed by \citet{gib2003} to include more complex structures, in particular streamers at mid-latitude. Uninterrupted observations of the solar corona over the course of half a rotation or longer has made solar rotational tomography possible \citep[for example]{dav1994, zid1999, fra2000}. These tomographic reconstructions have concentrated on the solar minimum corona. A detailed 3D reconstruction of density within a streamer sheet at heights above 2.5\Rs\ is made by \citet{the2006}, where the face- and edge-on appearance of an isolated high-latitude streamer with solar rotation is used to constrain both the electron density as a function of height, and the variation of density across the streamer sheet. All these empirical models give a valuable measurement of density, and the 3D distribution of density in the corona, but cannot, or are not directly aimed at, resolving the link between structures in the corona and features on the solar disk.

% A number of models of large-scale magnetic coronal structure based on potential field source-surface (PFSS) extrapolations of photospheric magnetic field data \citep{alt1969}, or of 3D magnetohydrodynamic (MHD) models REFERENCE!!!\citep{}, which achieve a steady-state solution of the coronal magnetic field based on initial conditions dictated by observations of the photospheric field. The prediction of the position of the heliospheric current sheet given by PFSS models is in good agreement with observations of the streamer belt at solar minimum \citep{wang1992}, and gives a reasonable agreement outside of solar minimum. Global MHD models give a more detailed model of the magnetic field configuration at low heights in the corona. Given the great observational advances of the past decade, and in anticipation of future missions such as the Solar Terrestial Relations Observatory (STEREO) mission \citep{how2002}, it is increasingly feasible to build density models based directly on coronal observations. The structure within such models can serve as a proxy for the coronal magnetic field, and can be complimentary to the results of PFSS or global MHD models. 

Streamers are structures in the corona with densities around 3 to 10 times higher than surrounding regions \citep[for example]{sch1953,woo1999}. They are large and often long-lived structures which extend from the base of the corona out into the interplanetary solar wind \citep{jing2002}. Their sources on the solar surface are areas of enhanced magnetic activity - active regions associated with sunspots or prominences \citep{lou1989,kou1992}. Prominences (or filaments when viewed against the solar disk) are cool, high-density ribbons of material which lie in the hot and tenuous corona. The association of large helmet streamers with prominences in the low solar atmosphere has, for many decades, been made from eclipse observations (see \citet{sai1973}, and references within). \citet{bel2004} have shown, from a whole solar cycle analysis of the white light and the 530.3 nm emission corona, that the distribution of helmet streamers migrates with the distribution of prominences. Perhaps contradictory, \citet{liewer2001} studied 8 major streamers during April 1998, and found a strong correlation between their alignment and the position of active regions on the disk. 

In this work, we build an empirical 3D model of the large-scale structure of the corona. The model is based on the assumption that high-density sheets arise from filaments (or prominences) in the chromosphere. Constraints on the position of these sheets are imposed by disk observations of H$\alpha$ filaments. These sheets are forced, with increasing height, to twist and merge, and form the stalks of streamers as observed by the LASCO C2 coronagraph at heights above where the corona becomes radial (or very close to radial), providing a constraint on their position in the extended corona. The agreement between the modeled and observed coronal images, and the simulation of the changing appearance of the corona with solar rotation, is then a test of the basic assumptions used to build the model. 

%All observed and synthesized images are processed with a normalizing radial-graded filter (NRGF), described by \citet{morgan2006}. The NRGF calculates the average and standard deviation of brightness as a function of height within an image, and uses these values to normalize the brightness at each height. Therefore viewing the coronal structure seen in a NRGF-processed image is equivalent to viewing a large set of normalized latitudinal profiles simultaneously. Since, in using the NRGF, we are comparing synthesized and observed sets of normalized latitudinal profiles, the exact radial profile of density becomes a secondary concern. This allows us to focus on the latitudinal profile of density, and gives a qualitative comparison of synthesized and observed images. This is similar to the approach of \citet{wang1997}, who in building a 3D density model assume an arbitrary radial dependence of density of $r^{-2.5}$ within a narrow region surrounding the heliospheric current sheet.

Observations showing the large-scale coronal structure during the end of CR 2005, and the start of CR 2006 are shown in section \ref{observations}. The density model is described in section \ref{model} and the images calculated from the model are compared to observations in section \ref{results}. The merits of the model in exploring structure, suggestions for improvements, and the implications of the results are discussed in section \ref{discussion}. Conclusions are given in section \ref{conclusions}.

\section{Observations}
\label{observations}
\begin{figure}
\centering
\includegraphics[width=8.5cm]{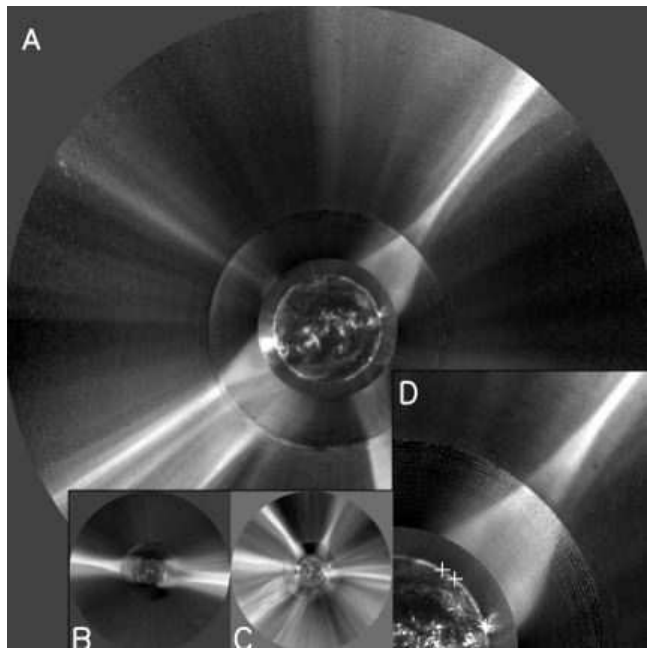}
\caption{Composite images of the corona made during (\textbf{A}) 1 August 2003, (\textbf{B}) 18 January 1997 (solar minimum), and (\textbf{C}) 8 December 2000 (solar maximum). The innermost views of the disk and low corona are EIT images of the Fe IX/X 171\AA\ line, except for the solar minimum disk which shows the He II 304\AA\ line. The corona from 1.15 to 2.3 $ R_{\odot}$\ are from the MLSO coronameters and above 2.3 $ R_{\odot}$\ from the LASCO C2 coronagraph. (\textbf{D}) Detail of the large north-west helmet streamer seen in \textbf{A}. The position of prominences and an active region at the limb are shown as crosses.}
\label{image}
\end{figure}

Figure \ref{image}A shows the structure of the corona on 2003/08/08 19:00. The image is composed of the SOHO Extreme Ultraviolet Imaging Telescope (EIT) observations of the Extreme UltraViolet (EUV) Fe IX/X 171\AA\ lines \citep{del1995}, and polarized brightness (\pB) observations by the Mauna Loa Solar Observatory's (MLSO) MK IV coronameter and the LASCO C2 coronagraph (see \citet{fisher1981} for a description of the MLSO MKIII coronameter). At heights above the limb, the images have been processed using a Normalizing Radial-Graded Filter (NRGF), described by \citet{morgan2006}. The NRGF calculates the average and standard deviation of brightness as a function of height within an image, and uses these values to normalize the brightness at each height. Therefore viewing the coronal structure seen in a NRGF-processed image is equivalent to viewing a large set of normalized latitudinal profiles simultaneously. Images of the corona made near the minimum and maximum of activity shown in Figs. \ref{image}B and \ref{image}C show that the general large-scale structure of the corona during 2003 was neither as complex as during solar maximum, nor as simple as during solar minimum when streamers appear primarily at the equator. Choosing a relatively simple corona with isolated streamers at mid- to high-latitudes greatly facilitates the analysis of the large-scale coronal structure. Figure \ref{image}D shows details of the large helmet streamer seen on the north-west limb in figure \ref{image}A, in relation to the position of two prominences (the two northernmost white crosses) and an active region/prominence complex (white cross near the equator) at the west limb, from which the legs of the streamer seem to arise.

\begin{figure} 
\centering \includegraphics[width=9cm]{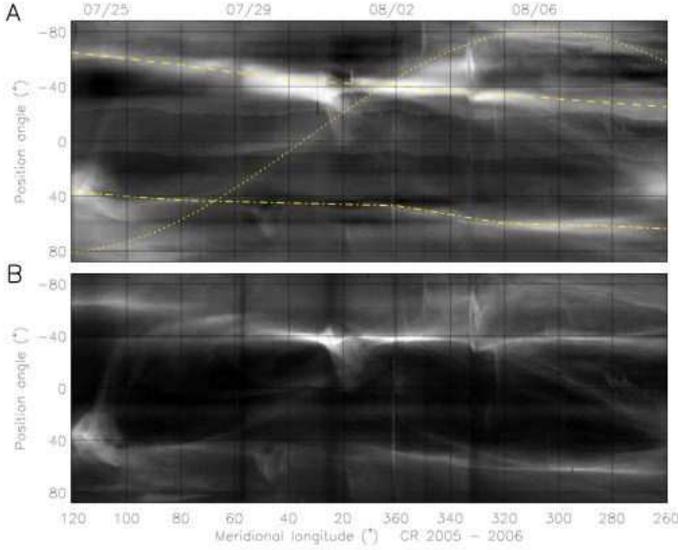}
\caption{Synoptic maps of white light brightness created from LASCO C2 latitudinal profiles stacked in time for heights of (\textbf{A}) 2.5$ R_{\odot}$\ and (\textbf{B}) 4.2$ R_{\odot}$. The upper $x$ axis gives dates (during 2006) and the lower $x$ axis shows the corresponding meridional Carrington longitude. Note that time goes left to right. The $y$ axis shows the position angle measured from north (positive is on the east limb, negative is west). The yellow paths in (\textbf{A}) trace the positions of three structures described in the text.}
\label{carr}
\end{figure}

During the period from 2003/07/25 to 2003/08/10, LASCO C2 made over 800 total brightness observations. This set of images is calibrated using the standard procedures of the LASCO Solar Software package, and estimates of the F-coronal brightness (the dust contribution) and stray light brightness are subtracted using procedures described in \citet{morgan2006}. Latitudinal profiles are extracted from the images at various heights. Each profile is normalized to a mean of zero and a standard deviation of one. The profiles are then stacked in time to give the final maps, shown in figure \ref{carr} for heights of 2.5 and 4.2\Rs. As shown by the dotted yellow line in Fig. \ref{carr}A, one streamer traces a quasi-sinusoidal curve with a period of a solar rotation. The phase and amplitude of this curve associates this streamer with a small active region just north of the equator near longitude 10-30$^{\circ}$, seen in the disk synoptic map of Fig. \ref{sourcestrmr}A. The most prominent structures in Fig. \ref{carr}, however, are the two streamers highlighted by dashed and dot-dashed lines at the west and east limb respectively. These drift slowly in latitude and do not describe the quasi-sinusoidal path of an isolated bright structure rotating with the Sun. During the two weeks, the west-limb streamer migrates slowly in the plane of the sky from -60$^{\circ}$\ to -20$^{\circ}$\, while the east limb streamer varies between 30$^{\circ}$\ and 65$^{\circ}$. To maintain such a slow variation, the structure of the north corona is likely to be dominated by an extended longitudinal band of high density. The synoptic maps shown for two heights in figure \ref{carr} are very similar, a result of the predominantly radial large-scale corona at this phase of the solar cycle. However, structures narrow considerably between 2.5 and 4.2$ R_{\odot}$, where the brightest central regions of the streamers are confined to very narrow latitudinal bands, generally less than 5$^{\circ}$. This is typical of large helmet streamers, and has been shown for solar minimum equatorial streamers by  \citet{guh1996}, \citet{wang1997} and others.

\section{Modeling the Coronal Structure}
\label{model}

We aim to model the large-scale structure of the north corona for dates 2003/08/01 to 2003/08/08. To achieve this, we create a 3D grid within which we can place structures of varying shapes and positions. The shape and positions can be adjusted as a function of height. The grid can be rotated to simulate solar rotation, and synthetic images may be created for comparison with observations by integrating along the various lines of sight (LOS). As will be described, observations are used to constrain the position and shapes of streamers at the solar surface and at a height of 4.2\Rs. Above this height, streamers maintain a radial structure. Between these heights, the streamers evolve smoothly from their initial to their final configuration.

\begin{figure} 
\centering \includegraphics[width=9cm]{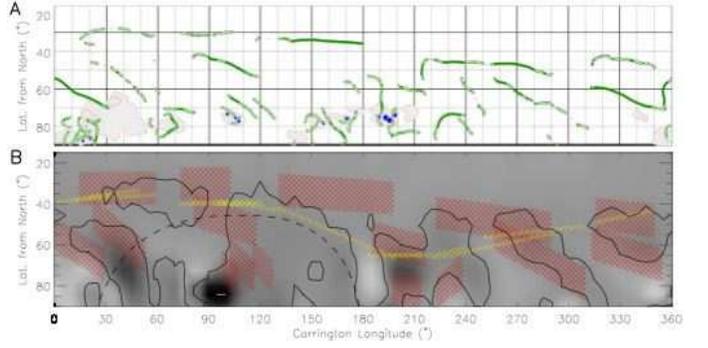}\caption{(\textbf{A}) Paris-Meudon Observatory (PMO) synoptic map  of Carrington rotations 2005 and 2006 showing prominences (green lines), sunspots (blue spots) and plage (grey areas). (\textbf{B}) Position of modeled structures at the solar surface (red areas) and above 4.2$ R_{\odot}$\ (yellow areas). The background image gives the photospheric magnetic field measured by the Wilcox Solar Observatory (WSO). The solid contour gives the photospheric magnetic neutral line and the dashed contour gives the coronal magnetic neutral line, calculated by PFSS extrapolation of the photospheric magnetic field for a height of 2.5$ R_{\odot}$.}
\label{sourcestrmr}
\end{figure}

The positions of the footpoints of high-density structures at the solar surface are constrained by the positions of filaments, as given by the synoptic map created by the Paris-Meudon Observatory (PMO) shown in figure \ref{sourcestrmr}A. The filaments are observed in H$\alpha$ near the central meridian, and are drawn in the synoptic maps as green lines. A set of latitude-longitude coordinates along each of these filaments is recorded, and areas of the model corona at the solar surface within $\pm$8\de\ of a filament, are given a high density. These are shown as the red cross-hatched regions in figure \ref{sourcestrmr}B. We do not correct for differential rotation in creating our density model, and accept that some filaments may change shape and position between the time of their observation at the meridian, and their passing at the east or west limb. 

The white-light synoptic map shown for a height of 4.2\Rs\ in figure \ref{carr}, provides the constraint for the coronal structure at this height, and above. The latitude and thickness of the high-density structures are given by the latitude and approximate thickness of the streamers at the appropriate longitude in the synoptic map. The narrow yellow cross-hatched regions in figure \ref{sourcestrmr}B show the positions and widths of these structures at a height of 4.2\Rs. For a given date, we assume the position of the bright streamer stalk given by the white light synoptic map as the true position of the streamer stalk. This is not strictly correct at all times due to line of sight issues, since the position angle of a streamer seen in a white light image can only give a true latitude if we know that the streamer stalk is directly above the limb. However, this approximation is a reasonable one for the north corona during the time in consideration, since figure \ref{carr} shows a slow migration of streamer stalks with latitude over the course of weeks.

For the sake of computational efficiency, not all filaments observed on the disk are included in the model. For example, we have not included smaller filaments which are far from the limb near the times in consideration since they have only a minimal impact on the appearance of the corona at these times. Note also that the PMO synoptic map shown in figure \ref{carr} contains regions of the disk from the end of Carrington rotation 2005 (2003/07/6 - 2003/08/2) and the start of rotation 2006 (2003/08/2 - 2003/08/29), to minimize the time between the observation of the filament at the meridian, and its passing at the east or west limbs.

\begin{figure} 
\centering \includegraphics[width=9cm]{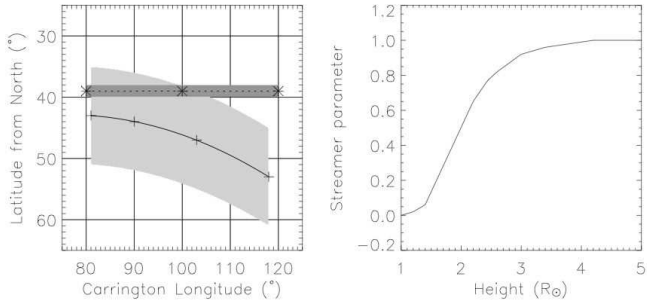}\caption{\textit{Left} - Example showing the modeled configuration of a single high-density sheet at the solar surface (lightly shaded region) and at 4.2\Rs\ (darkly shaded region). The crosses are a set of coordinates describing the position of the filament at the solar surface. The stars show coordinates describing the position of the streamer stalk in the corona. Both sets of coordinates are well constrained by observation. The solid line is a polynomial fit to the surface coordinates, and the dotted line a fit to the coronal coordinates. \textit{Right} - $f(r)$, the function of height used to describe the evolution of streamer parameters from the surface to the corona. This function is determined empirically by comparison of the shape of the modeled and observed streamers. All structural parameters describing the high-density sheets (position, shape and thickness) evolve with height following this curve.}
\label{curve}
\end{figure}

An important element for creating the model corona is the evolution of the positions and shapes of the density structures between the solar surface (constrained by the observed positions of filaments) and a height of 4.2\Rs\ (constrained by the white-light synoptic maps). As an example, we describe in more detail the modeling of an individual high-density structure. The left plot of figure \ref{curve} shows the surface (lightly shaded region) and coronal (darkly shaded region) configuration of one high-density sheet. At the solar surface, a set of longitude ($\theta$) and latitude ($\phi$) coordinates (shown as crosses) describing the alignment of the filament on the disk is fitted to a 2-degree polynomial $\phi = c_0 + c_1 \theta + c_2 \theta^2$, shown as a solid line in figure \ref{curve}. At a height of 4.2\Rs, another set of coordinates (shown as stars) describe the position of the streamer as seen in a coronal synoptic map. These are also fitted to a 2-degree polynomial $\phi = c_0^\prime + c_1^\prime \theta + c_2 ^\prime \theta^2$, shown as a dotted line. We assume that the angular width ($\sim$40\de\ in this example, corresponding to the length of the filament) of the high-density structure remains the same between the surface and the corona (i.e. angular width is conserved with height in latitude-longitude space, meaning an actual increase in true width), and that the central longitude ($\sim$100\de) is maintained. The streamer changes position from the surface to the corona by adjusting the set of coefficients $c_i$ from their initial ($c_i$) to their final ($c_i^\prime$) values, so at height $r$ the latitude across a streamer is given by

\begin{equation}
\phi= c_0 + (c_0^\prime-c_0)f(r) + [(c_1^\prime-c_1)f(r) + c_1]\,\theta+ [(c_2^\prime-c_2)f(r) + c_2]\,\theta^2.
\end{equation}
\noindent
$f(r)$ is a function which increases from zero at the solar surface to unity at a height of 4.2\Rs. Thus each parameter describing the streamer adjusts with height from the initial value at the surface to the final value at 4.2\Rs, with the height evolution described by the function $f(r)$. Its form is determined empirically from the general shape and thickness of streamers in coronal images, and is shown in the right plot of figure \ref{curve}. $f(r)$ is steepest near 2\Rs. The gradient lessens above $\sim$2.5\Rs\ to enable a smooth transition into the final radial configuration. $f(r)$  changes little between 3\Rs\ and the final configuration at 4.2\Rs, therefore our model corona is very close to radial at 3\Rs. Another important parameter controlled by $f(r)$ is the thickness of the streamer, which decreases with height from 16\de\ at 1\Rs\ to 3\de\ at 4.2\Rs.

\section{Results}
\label{results}
Synthetic polarized brightness images are created from the coronal density model and are compared to observed images of the corona. We do not attempt to determine the true electron density of the background corona or streamers. The background density is set to a value typical of coronal holes within our model, using values given by \citet{doyle1999}. The density inside streamers is set to values found by \citet{guh1996} for solar minimum equatorial streamers. At a given height, the density within a streamer is uniform. This is similar to the approach of \citet{wang1997}, who in building a 3D density model assume an arbitrary radial dependence of density of $r^{-2.5}$ within a narrow region surrounding the heliospheric current sheet. The standard LOS integration for polarized brightness is used \citep{van1950,que2002}, and the tilt of the Sun's rotation axis relative to Earth is included. The radial gradient of brightness within the image is then removed using the NRGF. It is this step which allows a qualitative comparison of modeled and observed structures, which are also processed with the NRGF. Since, in using the NRGF, we are comparing synthesized and observed sets of normalized latitudinal profiles, the exact radial profile of density becomes a secondary concern. This allows us to focus on the latitudinal profile of structures, and gives a qualitative comparison of synthesized and observed images. This qualitative comparison is a valid but non-rigorous method to explore the large-scale coronal structure - one which greatly reduces the number of necessary model parameters, and allows us, for a preliminary study of structure, to neglect the derivation of an absolute electron density. 

\begin{figure} 
\centering \includegraphics[width=9cm]{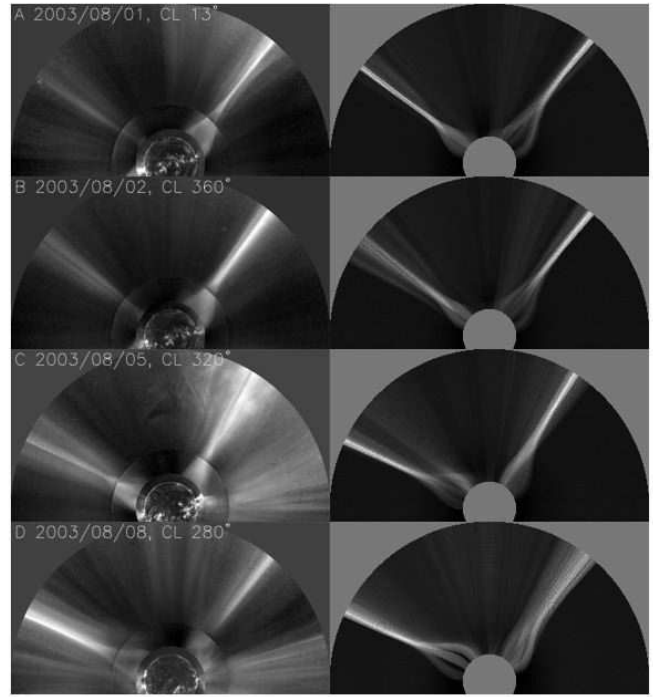}\caption{Comparison of observed coronal images (left column) and the synthetic images (right column). The observed coronal images are composites of EIT 171\AA, MLSO MKIV pB and LASCO C2 pB, all processed by the NRGF. The synthetic images show \pB\ calculated from the coronal density model, also processed by the NRGF. The four sets of observations were made within one to two hours of 19:00 for dates 2003/08/01, 2003/08/02, 2003/08/05 and 2003/08/08 (top to bottom). The dates and meridional Carrington longitude are given in the top left corner of each image.}
\label{datamodel}
\end{figure}

Figure \ref{datamodel} shows four observed and modeled pairs of images for four dates between 2003/08/01 and 2003/08/08. Good general agreement of large-scale structure is achieved in all four images. At large heights, the apparent latitudes and widths of the 2 main streamers are respected. In the first two pairs of images, for meridional Carrington rotation 13\de\ and 360\de, the general shape of the large helmet streamer on the west limb is well replicated. The sources of this streamer are the prominences widely separated in latitude at Carrington longitude $\sim$80-120\de, as shown in figure \ref{sourcestrmr}. To replicate the shape of the observed helmet streamer, the individual modeled high-density sheets arising from each prominence merge into one narrow band at latitude 40\de. The changing shape of this streamer through all four dates is also well reproduced. Between rotation 360\de\ and 320\de, this streamer narrows at the base, and by rotation 280\de, the classic helmet streamer shape is lost. The prominences on the west limb by this time, near a longitude of 30\de\ in figure \ref{sourcestrmr}, are less separated in latitude. Also the streamer sheet at a height of 4.2\Rs\ is inclined along the line of sight, and appears less narrow and bright.

For the first two dates, the main east-limb streamer is too broad at low heights in the model. There may be several reasons for this - that the modeled prominences are too broad (see longitudes 250\de-290\de\ in figure \ref{sourcestrmr}), the merging of the structures from the two sources should happen at a lower height (the form of $f(r)$ may be inappropriate for this streamer), or, more basically, the assumption that high-density sheets arise from disk filaments is incorrect in this case.  For the final two pairs of images, at rotation 320\de\ and 280\de, the helmet shape of the east-limb streamer is well reproduced. The shape of this streamer is also caused by widely separated sources merging into one narrow structure (see longitudes 180\de-240\de\ in figure \ref{sourcestrmr}).

\section{Discussion}
\label{discussion}

It is not surprising that the large-scale coronal structure is well replicated at heights above $\sim$3\Rs\ in the synthesized images of figure \ref{datamodel}, since we have constrained the position of our high-density sheets by the apparent latitude of observed streamer stalks in LASCO C2 synoptic maps. The main interest, or purpose, of the modeling lies in the agreement at heights below 3\Rs. From the reasonable agreement between observed and synthesized images, there is no doubt that filaments in the chromosphere (or the magnetic field reversals traced by, or lying above, these filaments) have an important role in defining the large-scale structure of the north corona during the period under consideration, and that the bulk of streamer activity is not directly associated with active regions. An interesting case is the large helmet streamer observed on the west limb on 2003/08/01. The longitude at the west limb is $\sim$100\de\ on this date, and from the PMO synoptic map of figure \ref{sourcestrmr}A, we see that the base of the helmet streamer encompasses two mid-latitude filaments, and an active region/filament complex nearer to the equator. The positions of these disk features is shown in relation to the helmet streamer in figure \ref{image}D, and in more detail in the left image of figure \ref{latprof}. It seems, from the alignment of the bright (hot) emission seen both in the EIT 171\AA\ part of figure \ref{image}D, and in the EIT 304\AA\ part of figure \ref{latprof}, that both active regions and prominences play a role in the formation of this streamer.

\subsection{Helmet streamer structure}
\begin{figure} 
\centering \includegraphics[width=9cm]{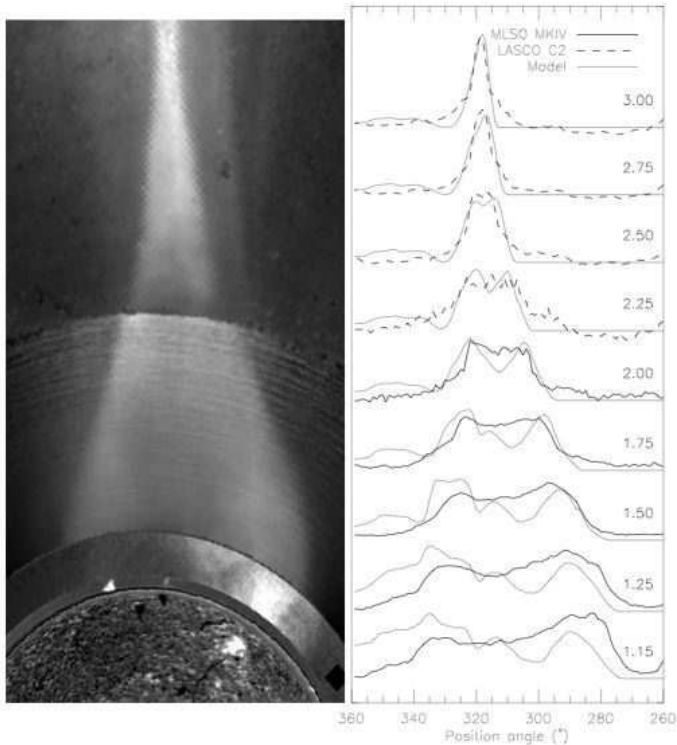}\caption{\textit{Left} - Composite image of the large helmet streamer seen on the north west limb during 2003/08/01, observed by Advanced Coronal Observing System's (ACOS) Chromospheric Helium Imaging Photometer (CHIP) in He I 10830\AA\ intensity (0-1.06\Rs), EIT He II 304\AA\ (1.06-1.25\Rs), MLSO MKIV coronameter (1.25-2.2\Rs) and LASCO C2 (above 2.2\Rs). This image is rotated so that the streamer is aligned approximately vertically, the true position angle of the streamer stalk being $\sim$318\de. \textit{Right} - Latitudinal brightness profiles across the large helmet streamer. Profiles are given for the MLSO MKIV coronameter's daily average of \pB\ (solid bold lines), LASCO C2 \pB\ (dashed bold lines) and \pB\ calculated from the density model (lighter solid lines). Heights are given in \Rs\ to the right of each set of profiles. For clarity, all profiles are normalized by their standard deviation, and are staggered to separate profiles at different heights. Position angle is measured counterclockwise from North.}
\label{latprof}
\end{figure}

While the general shape of streamers is well simulated by our model, a shortcoming lies in the discrepancies seen within the internal structure of streamers at low heights. Taking the large north-west helmet streamer of figure \ref{datamodel}A as an example, we show a detailed image of this streamer in the left image of figure \ref{latprof}. We extract latitudinal profiles of \pB\ at various heights across the streamer, from model and observation, and compare them in the right plot of figure \ref{latprof}. At the lowest heights (1.15-1.75\Rs), the model profiles have three distinct peaks within the large streamer, corresponding to the three sets of H$\alpha$ filaments seen in the synoptic map of \ref{sourcestrmr}A at longitude $\sim$100\de\ on this date, and which we have included in the model as the positions of high-density sheets. The observed profiles, however, do not contain a central peak. By 2\Rs, this central peak in model brightness has merged with the northernmost leg of the streamer. The model streamer now has two distinct peaks at each edge, with a dark void in the core (although the brightness of this void is considerably higher than the background). This is an exaggeration of what we see in the observations, where there is only a slight dip in brightness across the streamer. At the lowest height profile given for LASCO C2 (2.25\Rs), the same is true - the model streamer is still composed of two high-density sheets which have still not merged, and the space between them leads to a void which is too dark compared to observation.  It is possible that these failings could be improved by the details of the modeling. Some obvious improvements would be:

\begin{itemize}
\item Filaments (or filament channels) are narrow structures observed in H$\alpha$, increasing with height above the chromosphere \citep{engvold1976}. We have rather arbitrarily modeled high-density sheets with an angular width of $\pm$8\de\ centered on filaments. \citet{fou2006} present new methods for detecting prominences in EIT images. The synoptic maps they create show prominence regions covering very wide areas of the low solar atmosphere, with varying widths, but often of the same order, or even broader, as those used in this work. It will be interesting to use these EIT maps to more accurately map our high-density sheets in the low corona. This may well help in providing extra brightness within the bases of large helmet streamers, and in correcting other discrepancies such as the overly broad base of the modeled northeast-limb streamer seen in figures \ref{datamodel}A and B. An additional improvement gained from using EIT synoptic maps would be the direct observations of prominences on the limb, thus reducing an unnecessary uncertainty involved in this current work, which arises from the $\sim$7 day interval between observations of filaments at the meridian and their passing of the east or west limbs. 
\item The final appearance of the model corona at low heights is very sensitive to the chosen form of the function which controls the parameters describing the evolution of structures with height ($f(r)$, shown in figure \ref{curve}). In this work, one $f(r)$ is used to control all streamers. Determining a separate $f(r)$ for each streamer, or high-density sheets, would allow far greater control of the shape of that streamer, and would guarantee a better fit with observation.
\item Flattening the images with the NRGF prior to comparing model with observation lessens the importance of absolute density as a function of height, but the relative density of the streamers is an important factor which is not included, and of course has a direct impact on their relative brightness in images.
\item Also important is the variation of density across a streamer sheet. This can vary considerably, possibly by a factor of 10 \citep{the2006}. Our modeled streamers have a uniform density at a given height.
\item We have not considered the role of high-density structures emerging from active regions. Although active regions are restricted to latitudes close to the equator during this period (see figure \ref{sourcestrmr}), they do have a contribution to the coronal brightness at higher latitudes, forming parts of large helmet streamers, as can be clearly seen in figure \ref{latprof}, or, if they are isolated structures, as they transverse across the pole with solar rotation \citep{jing2000}. 
\end{itemize}

\noindent
Without implementing these improvements, it is difficult to say whether the cores of the modeled helmet streamers would agree better with observation, or if there is a flaw, or a missing element, in the structural assumptions used to create the model.

\begin{figure} 
\centering \includegraphics[width=8.5cm]{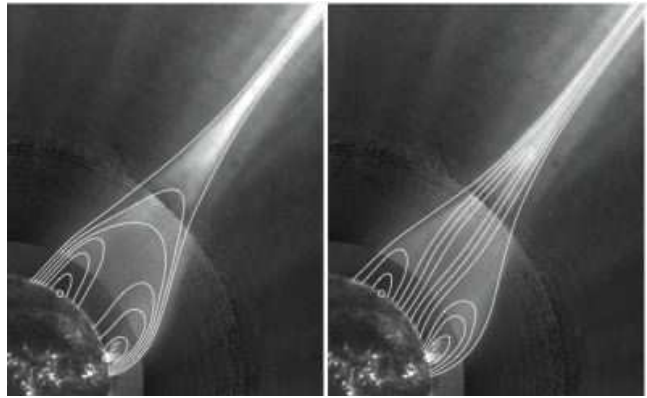}\caption{Schematic illustrations of possible magnetic topologies within large helmet streamers. \textit{Left} - the interior cavity of the helmet streamer is composed exclusively of large arches of closed field while open field originates from surrounding regions. \textit{Right} - the closed field dominates only small regions directly associated with the prominences and/or active regions. Open field can originate between the closed field structues, merging into the helmet streamer, as well as from neighboring regions.}
\label{schematic}
\end{figure}

Creating a model of the corona based on high-density sheets, with extended longitudes above the observed position of filaments, has given us a good overall agreement of the outline of large helmet streamers, in other words, the legs of helmet streamer are well simulated. This is in support of the multiple current sheet model of helmet streamers \citep{eddy1973,crooker1993}. Indeed, without resorting to an empirical model, the direct observation of enhanced brightness in the streamer legs, correlated with the position of prominences/active regions in the low corona is supportive of this view. Figure \ref{schematic} illustrates schematically two possible views of the magnetic topology within the large helmet streamer. The left schematic is based on a large system of closed loops which form the streamer core. In the right schematic, systems of closed loops are smaller, and localized directly above prominences or active regions. The helmet shape of the streamer is caused by the merging of open field over a large region, and the overall topology of the streamer is one of multiple current sheets. The structure of our model is based on the latter view, and it has considerable success in replicating both the overall shape of streamers, and the sheets of enhanced brightness (the streamer legs) directly above the position of prominences or active regions. On the other hand, the PMO synoptic map of figure \ref{sourcestrmr} shows one filament situated just North of the center of the streamer base. This prominence can be seen in the He I 10830\AA\ CHIP image of figure \ref{latprof}, underlying a small, dark, arched cavity extending above its position to a height of 1.4\Rs\ in the EIT 304\AA\ and the lowest height of the MKIV coronameter's field of view. This is very different from the enhanced brightness above the neighboring prominence, under the northernmost streamer leg. Is it possible then that some prominences have high-density sheets extending into the corona to form streamers, or the legs of large helmet streamers, while other prominences are surrounded by a larger system of closed loops? It is known that prominences can vary considerably in structure \citep[and references within]{heinzel2006}. If the closed field or arcade associated with a prominence is restricted to low heights, the current sheet forming above this arcade begins in a region with higher density, and other different plasma characteristics, than current sheets which form at larger heights above larger systems of closed field. \citet{merenda2006} found observational evidence for strong, near-vertical magnetic fields within a polar crown prominence, whereas previous studies showed that low- to mid-latitude prominences are dominated by a weaker horizontal field \citep{leroy1984,bommier1994}. Could vertical fields in high-latitude prominences form high-density sheets which are the legs of helmet streamers?

The enhanced brightness in the legs of the streamer shown in figure \ref{latprof} (the same can be seen also in the northeast helmet streamer of figure \ref{datamodel}D), is reminiscent of the profile of O VI intensity across some large helmet streamers \citep{noci1997a,ray1997a,frazin2003}, as observed by the UltraViolet Coronagraph Spectrometer/SOHO (UVCS). \citet{noci1997b} provides a multiple current sheet model of helmet streamers to interpret the UVCS observations. Unfortunately, we do not have dedicated UVCS observations of the large helmet streamers discussed in this work. That only some helmet streamers show a depletion of O$^{5+}$ in their cores is discussed by \citet{uzzo2006} in terms of the alignment and complexity of filaments forming the streamer base. Our structural density model would be useful to confirm this interpretation.

\subsection{Large-scale coronal structure}

Fig. \ref{sourcestrmr}B shows the final position of the streamers at 4.2$ R_{\odot}$\ as a narrow yellow band. This band extends around all longitudes, varying slowly in latitude from 30$^{\circ}$-70$^{\circ}$. In the model corona, this band of high-density forms, approximately, an inverted cone above the North pole. As we form the synthetic images of figure \ref{datamodel}, bright streamer stalks form at the north-east and north-west limbs, where sections of the cone are seen edge-on. These sections are also bright due to the weighting towards the plane-of-sky - due to both the drop in density with height, and the geometrical formulations of the line of sight calculation of \pB. Other sections of the cone (for example, situated above the North pole) are fainter, due to being seen more face-on, and being farther from the plane-of-sky. In the model images, latitudinal variations can be seen across these faint regions. This is due to the actual distribution of the high-density sheets around the cone. Another important factor is that the line of sight passes through structures at both the near and far sides of the cone. The faint structure we see above the north pole is qualitatively very similar to the type of structure we see in the same regions of the LASCO images. The agreement could be greatly improved by adjusting the relative density of the modeled structures.

\begin{figure} 
\centering \includegraphics[width=8.5cm]{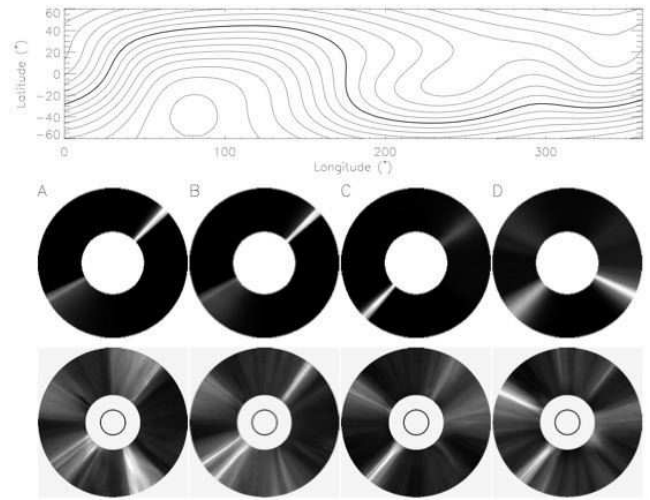}\caption{\textit{Top} - The current sheet predicted in the corona by PFSS extrapolation of WSO photospheric field measurements, for CR 2005 (the current sheet for CR 2006 is very similar). The extrapolation is made assuming a radial field at 2.5\Rs. The neutral line is given by the bold contour. Contours to the north of this line are negative, and to the south are positive, and give increments of the field in 1.5$\mu$T. The four columns of coronal images are shown for a time close to 19:00 for four dates - (A) 2003/08/01, (B) 2003/08/02, (C) 2003/08/05 and (D) 2003/08/08, corresponding to the observed and modeled pairs of figure \ref{datamodel}. The top row of images shows \pB\ calculated from a coronal density model, with a streamer density placed uniformly within $\pm$3\de\ of the PFSS current sheet. The bottom row of images shows LASCO C2 observations. All images are processed by the NRGF.}
\label{sheet}
\end{figure}

Shown as a dashed bold line in Fig. \ref{sourcestrmr}B is a section of the neutral line in the north hemisphere determined by a potential field source-surface (PFSS) extrapolation of photospheric magnetic field data. Only between longitudes 60$^{\circ}$\ and 160$^{\circ}$\ is there reasonable agreement between the magnetic field extrapolation and the position of the high-density band used in our model. This is an unsatisfying result, since the heliospheric current sheet calculated from PFSS has been shown to be an excellent predictor of the large-scale distribution of streamers at larger heights. Despite this success, it is hard to reconcile the appearance of the corona, during the period under consideration in this work, with the single current sheet model of the global coronal magnetic field calculated by PFSS extrapolation. This is demonstrated in figure \ref{sheet}, where a coronal current sheet given by PFSS extrapolation of WSO photospheric observations is used to create a density model of the corona. Regions in the corona within $\pm$3\de\ of the current sheet are given a high density typical of equatorial current sheets at solar minimum, as given by \citet{guh1996}. \pB\ is then obtained from the model by LOS integration at various stages of the solar rotation, corresponding to the dates studied in figure \ref{datamodel}. Using a uniform distribution of density, we see that PFSS is only successful in predicting the position of some of the largest and brightest streamers. The agreement would likely be much improved by introducing appropriate density variations across the heliospheric current sheet, and by considering rapid changes in the photospheric field \citep{wang2000}. Alternatively, a more complex configuration of current sheets can be considered \citep{saez2005}. \citet{liewer2001} considered that during times outside solar minimum, more filamentary structure, not predicted by the uniform heliospheric current sheet of PFSS models, is produced by newly opened field lines created by reconnection processes in active regions. Our model of the large-scale coronal structure, with development, can be seen as complementary to the PFSS calculation of the heliospheric current sheet, and may  give a better understanding of the longitudinal density variations along the current sheet which are necessary to better recreate the appearance of the corona.

%%%%%%%%%%%%%%%

\section{Conclusions}
\label{conclusions}

The model in this work contains an ensemble of high-density sheets directly associated with observed filaments on the disk.  Spreading the footpoints of streamers across broad regions centered on prominences, and allowing the merger of these structures into a narrow latitudinal band encircling the  corona as dictated by white light observations, proves to be a simple but effective way to reproduce the appearance of the corona, and the overall appearance of large helmet streamers. However, the model's replication of the internal or core structure of helmet streamer bases is unsatisfactory. This failing may be improved with stronger constraints on the footpoint configuration of high-density sheets at low heights (based possibly on EIT observations of prominences), and/or with a more realistic profile of absolute densities within structures. A more basic improvement is to broaden the scope of the model to include high-density structures from active regions. Additionally, our study of the large helmet streamer observed on the north-west limb on 2003/08/01 suggests that the model must have the flexibility to allow some prominences to posses high-density sheets which form streamers (or one leg of a helmet streamer) and allow others to form the bases of closed field regions or arches. The model, will, with such developments, serve to bridge a gap between models of coronal structure based on observations of the line-of-sight photospheric magnetic field (PFSS or global MHD models), and the increasingly detailed imaging observations of the Sun and corona.

The qualitative modeling of coronal structure made in this work can also be considered as a first step towards obtaining a quantitative measure of the coronal density. Indeed, once the qualitative 3D structure of the corona is established using the empirical model, electron density is easily calculated by inversion. This model provides a framework which can contribute to understanding the coexistence of closed and open magnetic field lines in the corona.

\begin{acknowledgements}
Our gratitude to Karen Teramura (IfA, Uni. of Hawaii) for preparing figure \ref{schematic}. The \pB\ data used in this work are courtesy of the MLSO and the LASCO/SOHO consortium. The MLSO coronagraphs and the ACOS CHIP instrument are operated by the High Altitude Observatory, a division of the National Center for Atmospheric Research, which is sponsored by the National Science Foundation (USA). SOHO is a mission of international cooperation between ESA and NASA. The Paris Meudon Observatory synoptic maps are available from \textit{http://bass2000.obspm.fr/}. The Wilcox Solar Observatory's coronal field extrapolations were obtained from the WSO section of Stanford University's website courtesy of J.T. Hoeksema. WSO is supported by NASA, the NSF and ONR.
\end{acknowledgements}

\end{document}